%% file: main.tex
\RequirePackage{amsmath}
\documentclass[runningheads]{llncs}
\usepackage[T1]{fontenc}
\usepackage{graphicx}
\usepackage{amssymb}
\usepackage{amsfonts}
\usepackage{xcolor}
\usepackage{subcaption}
\usepackage{multirow}
\usepackage[misc]{ifsym}
\usepackage{footmisc}

\newcommand{\ours}{\texttt{TC-BrainTF}}

\begin{document}

%\iffalse
\author{Yanting Yang\inst{1,2}\thanks{These authors contributed equally to this work.} \and Beidi Zhao\inst{1,2*} \and Zhuohao Ni\inst{1*} \and Yize Zhao\inst{3} \and \\ Xiaoxiao Li(\Letter)\inst{1,2}}
\authorrunning{Y. Yang et al.}
\institute{The University of British Columbia, Vancouver, BC V6T 1Z4, Canada \\ \email{xiaoxiao.li@ece.ubc.ca} 
\and Vector Institute, Toronto, ON M5G 1M1, Canada
\and Yale University, New Haven, CT 06520, USA}
%\else
\iffalse
\author{
Anonymous
}
\institute{Anonymous Organization \\
\email{**@******.***}
}
\fi

\title{Learnable Community-Aware Transformer for Brain Connectome Analysis \\ with Token Clustering}
\titlerunning{TC-BrainTF}
\maketitle

\begin{abstract}
Neuroscientific research has revealed that the complex brain network can be organized into distinct functional communities, each characterized by a cohesive group of regions of interest (ROIs) with strong interconnections. These communities play a crucial role in comprehending the functional organization of the brain and its implications for neurological conditions, including Autism Spectrum Disorder (ASD) and biological differences, such as in gender. Traditional models have been constrained by the necessity of predefined community clusters, limiting their flexibility and adaptability in deciphering the brain’s functional organization. Furthermore, these models were restricted by a fixed number of communities, hindering their ability to accurately represent the brain's dynamic nature. In this study, we present a token clustering brain transformer-based model (\ours{}) for joint community clustering and classification. Our approach proposes a novel token clustering (TC) module based on the transformer architecture, which utilizes learnable prompt tokens with orthogonal loss where each ROI embedding is projected onto the prompt embedding space, effectively clustering ROIs into communities and reducing the dimensions of the node representation via merging with communities. Our results demonstrate that our learnable community-aware model \ours{} offers improved accuracy in identifying ASD and classifying genders through rigorous testing on ABIDE and HCP datasets. Additionally, the qualitative analysis on \ours{} has demonstrated the effectiveness of the designed TC module and its relevance to neuroscience interpretations. 
% \keywords{First keyword  \and Second keyword \and Another keyword.}
\end{abstract}
{%
  \let\thefootnote\relax%
  \footnotetext{This work is supported in part by the Natural Sciences and Engineering Research Council of Canada (NSERC), Compute Canada and Vector Institute.}%
  \let\thefootnote\svthefootnote%
}

\section{Introduction}

The human brain is a vastly complex network of interconnected regions that facilitates cognition, perception, and behaviour~\cite{satterthwaite2015linked,bullmore2009complex}. Analyzing the functional connectivity patterns between these brain regions, known as the brain connectome, has emerged as a powerful tool for understanding neural mechanisms and diagnosing neurological and psychiatric disorders~\cite{kan2022bnt}. ROIs serve as fundamental units in connectome analysis, representing anatomically defined areas of the brain. Functional connectivity (FC) matrices, typically derived from correlations in functional MRI (fMRI) activity between ROI pairs, capture the strength of functional interactions across the brain network~\cite{smith2011network}. Analyzing these FC-based brain connectomes has yielded insights into cognitive processes and their disruptions in conditions like ASD~\cite{Kaiser2010,Price2014}.

In recent years, deep learning has significantly advanced the analysis of functional connectivity (FC) matrices, heralding new approaches to understanding brain networks. The BrainNetCNN~\cite{KAWAHARA20171038} innovated with convolutional filters specifically designed for the brain's complex network structures. Following this, Graph Neural Networks (GNNs)~\cite{veličković2018graph} embraced the brain's inherent non-Euclidean architecture, creating graphs from FC matrices to analyze connectivity. The evolution continued with the advent of transformer architectures~\cite{NIPS2017_3f5ee243}, renowned for their ability to model long-range dependencies between ROIs without relying on predefined graphs. A prime example of this is the Brain Network Transformer (BNT)\cite{kan2022bnt}, which utilizes transformer encoders to derive ROI embeddings from FC matrices and employs a readout layer for brain state prediction. More recently, new models have begun to recognize the significance of community-specific associations in the learning of ROI embeddings. For instance, community-aware transformer (Com-BrainTF)~\cite{bannadabhavi2023community} employs a hierarchical local-global transformer architecture to learn ROI embeddings that are aware of both intra and inter-community dynamics, showing improved performance and interpretability over previous arts. 
%These transformer-based brain connectome analysis methods have shown better performance on disease and age prediction tasks. However, 
However, Com-BrainTF heavily relies on predefined community structures on a given atlas\cite{bannadabhavi2023community,Craddock2012}. Neuroscience research has shown that brain network community patterns change across various tasks~\cite{betzel2019community,gadelkarim2014investigating}. This reliance on prior knowledge not only constrains the models' ability to generalize across diverse brain network configurations but also limits their flexibility and adaptability, ultimately compromising their capacity to capture the dynamic and intricate nature of the brain's functional organization\cite{bryce2021brain}.

To overcome these limitations in leveraging the inductive on brain regions functioning in groups in transformer design for brain connectome analysis, we improve the prior arts~\cite{kan2022bnt,bannadabhavi2023community} by enabling transformers learning to cluster brain regions. Following \cite{kan2022bnt,bannadabhavi2023community}, we treat each ROI as a token. Taking advantage of the tokenization and prompting mechanism, we introduce \ours{}, a learnable community-aware transformer that employs a novel token clustering strategy to dynamically learn community-specific ROI embeddings without necessitating predefined community for brain connectome. Central to \ours{} is a transformer encoder layer, a deep embedded clustering (DEC) layer with an orthogonal loss, and a readout layer, collectively enabling the flexible identification of community numbers and task-specific classifications. Our model surpasses traditional constraints by optimizing prompt tokens through orthogonal loss to enhance intra-community relations and using a soft assignment to project node features into clustered embedding spaces. The culmination of this process in a graph readout layer aggregates global-level ROI embeddings, providing a comprehensive brain graph representation. \ours{} not only allows for the discovery of previously unrecognized patterns, potentially offering new insights into disorders like ASD and gender differences but also ensures that the model's findings are driven by data characteristics, promoting objectivity and reproducibility. Validated through robust experiments, our model demonstrates a remarkable ability to discern functional community patterns essential for distinguishing between ASD and Healthy Control (HC), showcasing its potential to enhance our understanding of the brain's connectivity and contribute to neurological research.

\input{section/02-method}

\input{section/03-experiments}

\section{Conclusion}

In this work, we introduce \ours{}, a learnable, community-aware transformer architecture for brain network analysis with dynamic clustering using prompt tokens. Our model introduces a novel TC module with a set of learnable orthogonal prompts to cluster and merge ROI tokens. \ours{} not only outperforms SOTA baselines but also detects salient communities associated important for the classification task. We believe that this is the first work leveraging token clustering instead of pre-defined functional community information in transformer-based brain network analysis, providing a more flexible and powerful framework for brain connectome analysis. Our future work includes investigating better strategies to determine the optimal number of clusters $K$.
%Our framework is generalizable for the analysis of other neuroimaging modalities, ultimately benefiting neuroimaging research. Our future work includes investigating alternate variants for choosing different atlases and community network parcellations.

\bibliographystyle{splncs04}
\bibliography{main}

\clearpage
\newpage

\setcounter{equation}{0}
\setcounter{figure}{0}
\setcounter{table}{0}
\setcounter{page}{1}
\setcounter{section}{0}

\renewcommand{\theequation}{S\arabic{equation}}
\renewcommand{\thefigure}{S\arabic{figure}}
\renewcommand{\thesection}{S\arabic{section}}
\renewcommand{\thetable}{S\arabic{table}}
%\renewcommand{\thealgorithm}{S\arabic{algorithm}}

% \begin{table}[!t]
% \centering
% \caption{Comparing number of parameters.}\label{tab:para}
% \begin{tabular}{lr}
% \hline
% Method & \multicolumn{1}{l}{Number of parameters} \\ \hline
% FBNETGNN\cite{Fbnetgen} & 554694 \\
% BrainNetCNN\cite{KAWAHARA20171038} & 933277 \\
% BrainNetTF\cite{kan2022bnt} & 3772474 \\
% Com-BrainTF\cite{bannadabhavi2023community} & 8377042 \\
% \ours{} ($K=4$) & 10120738 \\
% \ours{} ($K=8$) & 10253858 \\
% \ours{} ($K=11$) & 10353698 \\ \hline
% \end{tabular}
% \end{table}

%\subsection{Ablation studies}

\begin{table}[!t]
\caption{We conduct the ablation study on the orthogonal loss. It can be seen that the addition of orthogonal loss improves the classification metrics in terms of AUROC, accuracy and specificity. For the classification task on the ABIDE dataset, the addition of orthogonal loss will improve the AUROC and accuracy by $2\sim3\%$, specificity by $16\%$. Even for the HCP dataset which the conventional transformer-based model has achieved relatively good results, the orthogonal loss can still improve the AUROC and accuracy by $0.3\sim0.4\%$. The drop in sensitivity reveals the potential limitation of applying orthogonal loss}\label{tab:ablation}
\centering
\resizebox{\textwidth}{!}{
\begin{tabular}{l|llll|llll}
\hline
\multirow{2}{*}{\begin{tabular}[c]{@{}l@{}}Orthogonal\\ Loss\end{tabular}} & \multicolumn{4}{c|}{ABIDE} & \multicolumn{4}{c}{HCP} \\ \cline{2-9} 
 & AUROC & Accuracy & Sensitivity & Specificity & AUROC & Accuracy & Sensitivity & Specificity \\ \hline
w/o & $76.2_{\pm 2.4}$ & $67.4_{\pm 3.6}$ & $\mathbf{74.4_{\pm 7.0}}$ & $60.3_{\pm 5.7}$ & $89.6_{\pm 3.0}$ & $80.6_{\pm 4.3}$ & $\mathbf{87.2_{\pm 2.8}}$ & $71.7_{\pm 9.4}$ \\
w/ & $\mathbf{77.7_{\pm 2.0}}$ & $\mathbf{69.4_{\pm 2.9}}$ & $69.1_{\pm 8.6}$ & $\mathbf{70.1_{\pm 3.9}}$ & $\mathbf{89.9_{\pm 2.1}}$ & $\mathbf{80.9_{\pm 2.5}}$ & $86.7_{\pm 3.8}$ & $\mathbf{73.1_{\pm 4.0}}$ \\ \hline
\end{tabular}
}
\end{table}

% We conduct the ablation study on the orthogonal loss.
% %and node representation in prompt embedding space that facilitates effective learning of node features and results in superior performance. 
% It can be seen from Table~\ref{tab:ablation} that the addition of orthogonal loss does not worsen the performance while improving the classification metrics in terms of AUROC, accuracy and specificity. %For the classification task on the ABIDE dataset, the addition of orthogonal loss will improve the AUROC and accuracy by $2\sim3\%$, specificity by $16\%$ while the sensitivity will drop by $7\%$. Even for the HCP dataset which the conventional transformer-based model has achieved relatively good results, the orthogonal loss can still improve the AUROC and accuracy by $0.3\sim0.4\%$, specificity by $2\%$, while the drop of sensitivity will only be $0.6\%$ compared with that on the ABIDE dataset.

\begin{figure}[!t]
\includegraphics[width=1.0\textwidth]{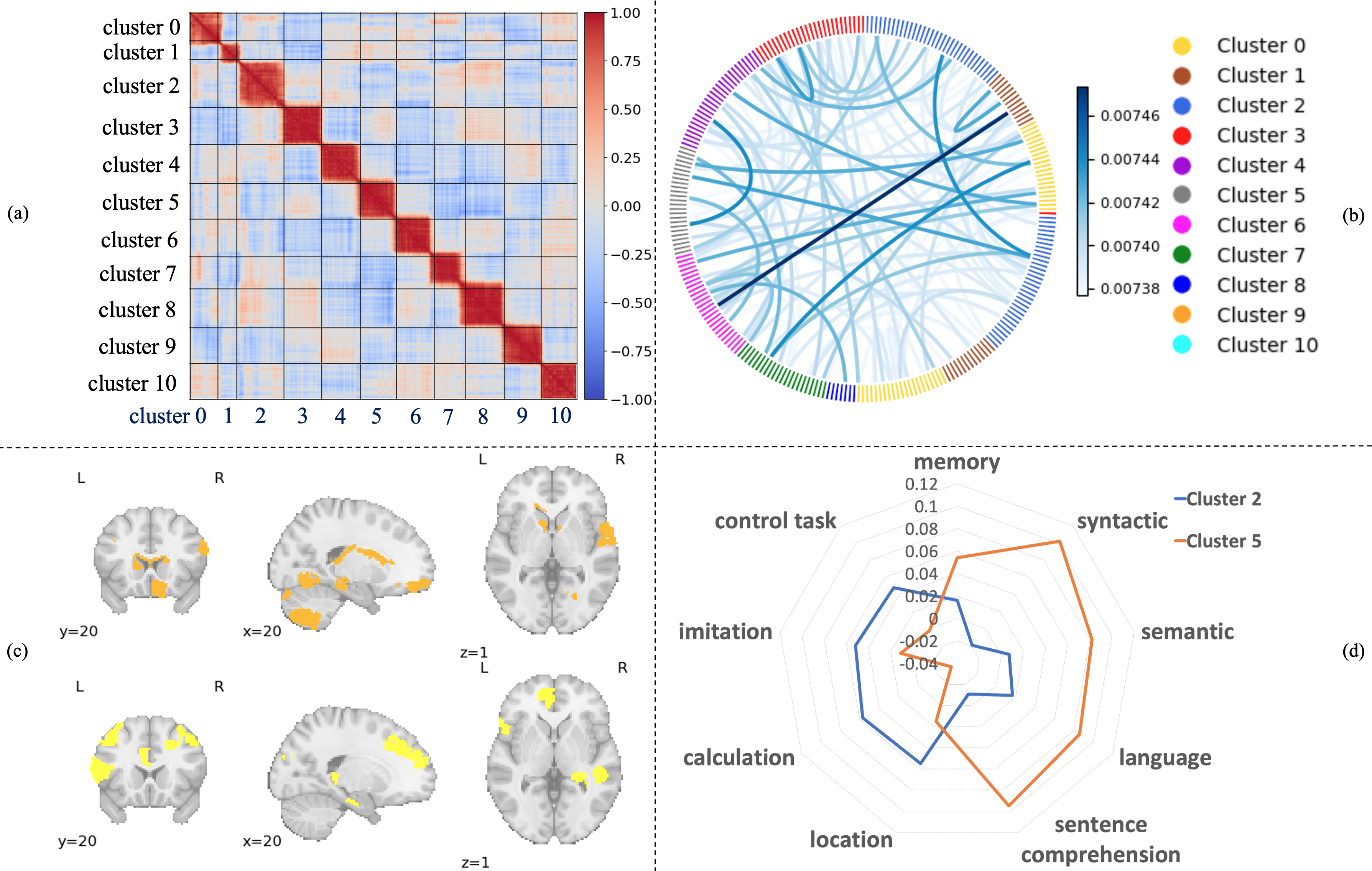}
\caption{Qualitative results for communities clustering of HCP dataset. (a) Assignment Matrix $A$ depicting the relationship between ROIs. (b) Chord diagram showcasing the top 0.5\% strongest connections from the token-token attention scores. (c) The ROIs in salient clusters 2 and 5. (d) Radar chart representing the correlation between functional clusters and cognitive keywords decoded by Neurosynth.}
\label{fig:HCP}
\end{figure}
\end{document}

%% file: section/02-method.tex
\section{Method}

\subsection{Overview}

\noindent\textbf{Data Representation.} The initial step parcellates the brain into $N$ Regions of Interest (ROIs) utilizing an atlas. From this, we construct a functional connectivity (FC) matrix for each subject by calculating the Pearson correlation coefficients between pairs of ROIs, reflecting the strength of fMRI co-activation. Consequently, for a brain graph consisting of $N$ nodes, we obtain a symmetric matrix $FC \in \mathbb{R}^{N \times N}$. In this matrix, columns represent the node length dimension, while rows indicate the node feature dimension, forming an $N$-length sequence with $N$-dimensional tokens. 

\iffalse
Upon identifying $K$ functional communities within the network of Regions of Interest (ROIs), our approach begins by initializing $K$ learnable prompt tokens, represented as $P \in \mathbb{R}^{K \times E}$, which act as preliminary community centroids in the embedding space. The prompt tokens and node feature embeddings are concatenated, forming a combined matrix $[P, X] \in \mathbb{R}^{(K+N) \times E}$. This matrix is then fed into a transformer encoder layer without position encodings, to produce $E$-dimensional tokens $H \in \mathbb{R}^{(K+N) \times E}$. This operation facilitates the learning of associations between ROI features and their corresponding community probabilities. A specialized deep embedded clustering layer then refines $H$ into embeddings $Z \in \mathbb{R}^{K \times E}$ and a soft assignment matrix $A \in \mathbb{R}^{N \times K}$, while computing an orthogonal loss $\mathcal{L}_\textrm{ortho}$. $Z$ symbolizes the node representation, $A$ embodies the soft assignment matrix detailing ROI community memberships, and $\mathcal{L}_\textrm{ortho}$ ensures the distinctness of community centroids. Finally, these node representations are interpreted by a readout layer to infer class probabilities, integrating learned embeddings and community assignments for predictive analysis.
\fi

\vspace{\baselineskip}

\noindent\textbf{Motivation and Pipeline.} The human brain connectome, marked by its hierarchical structure, demonstrates that ROIs within the same community share greater similarities than those across different communities. Traditional methods for delineating these communities and clustering nodes often rely on neurologist expertise and pre-established knowledge. To transcend these limitations, we designed a token clustering (TC) module composed of a deep embedded clustering (DEC) layer, an orthogonal loss, and a token merging strategy to dynamically learn community-specific node embeddings without predefined node clustering. This approach allows for the flexible identification of community numbers and provides task classifications along with a clustering assignment matrix. 

\begin{figure}[!t]
\centering
\includegraphics[width=.8\textwidth]{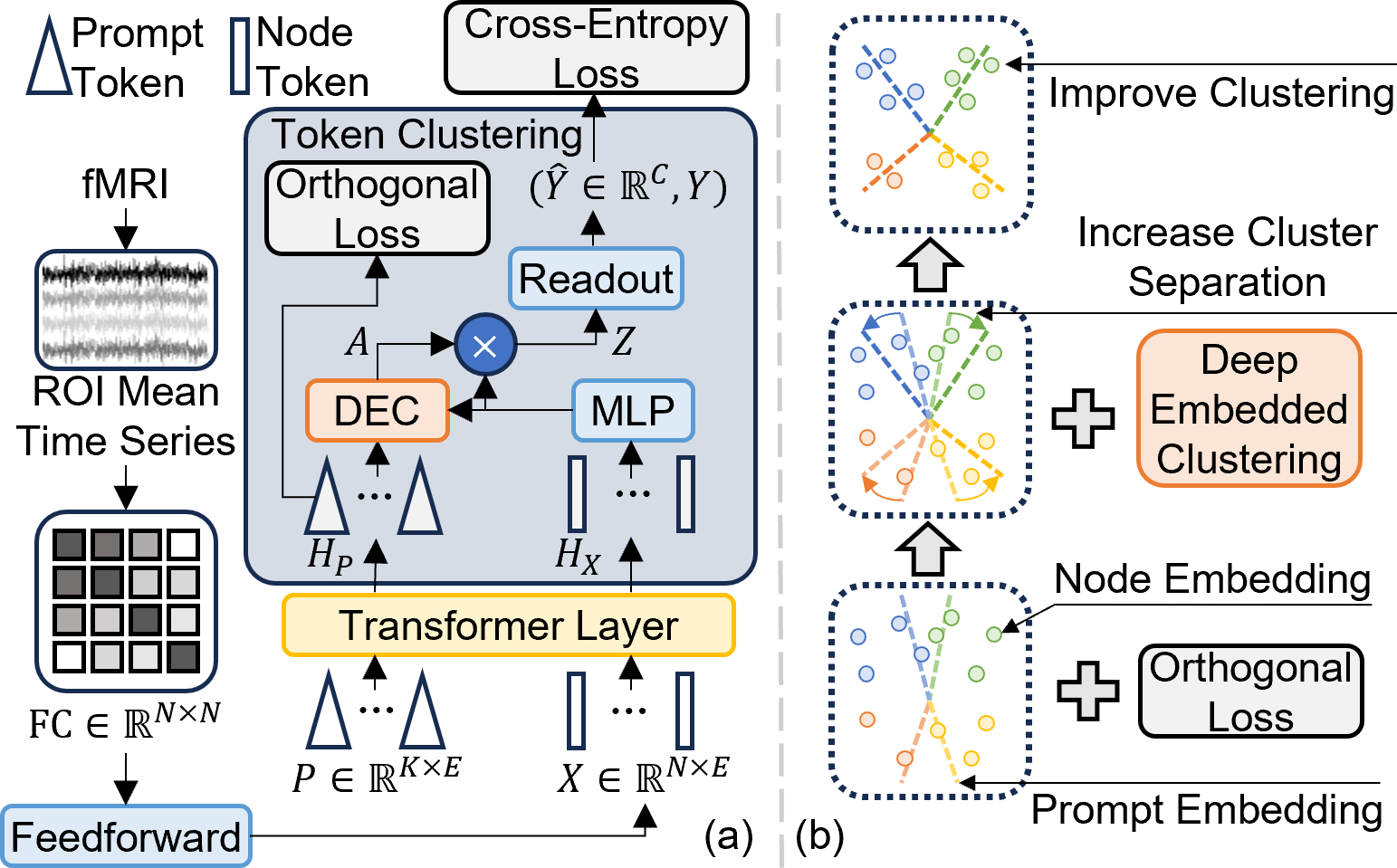}
\caption{(a) Overview of \ours{} framework. 
%$N$ selected ROIs from the fMRI scans are averaged along time and correlations are calculated between each ROI to form the FC matrix.
(b) Schematic of the orthogonal loss and orthogonal projection of node feature embeddings onto prompt token embeddings space.} \label{fig:architecture}
%\vspace{-6mm}
\end{figure}

Our model has three main components: a transformer encoder layer, a TC module and a readout layer. As shown in Fig.~\ref{fig:architecture}, the initial phase involves mapping this input $FC$ matrix to an embedding space $X \in \mathbb{R}^{N \times E}$ through a feedforward network. To identify $K$ functional communities using the DEC module, our approach optimizes $K$ learnable orthogonal prompt tokens, represented as $P \in \mathbb{R}^{K \times E}$, which act as preliminary community centroids in the embedding space. The prompt tokens and node feature embeddings are concatenated, forming a combined matrix $[P, X] \in \mathbb{R}^{(K+N) \times E}$. This matrix is then fed into a transformer encoder layer without position encodings, to produce $E$-dimensional tokens $H \in \mathbb{R}^{(K+N) \times E}$. This operation facilitates the learning of associations between ROI features and their corresponding community probabilities. A specialized deep embedded clustering layer then refines $H$ into embeddings $Z \in \mathbb{R}^{K \times E}$ and a soft assignment matrix $A \in \mathbb{R}^{N \times K}$, while computing an orthogonal loss $\mathcal{L}_\textrm{ortho}$. $Z$ symbolizes the node representation, $A$ embodies the soft assignment matrix detailing ROI community memberships, and $\mathcal{L}_\textrm{ortho}$ ensures the distinctness of community centroids. Finally, these node representations are interpreted by a readout layer to infer class probabilities, integrating learned embeddings and community assignments for predictive analysis. In the following subsection, we detail the operations in the key modules.

\subsection{Transformer Encoder Layer}\label{sec:TEL}

The transformer encoder~\cite{NIPS2017_3f5ee243} (Fig.~\ref{fig:architecture}(a)) is a crucial component of \ours{}. The transformer encoder takes $X$, the linear embedding of $FC$ as inputs (no position encodings). The transformer contains a multi-head-attention module to capture dependencies between nodes and prompt tokens using the attention mechanism. The learned token feature matrix $H$ is given by
\begin{equation}
H = \mathrm{concat}(\mathrm{head}_1, \cdots, \mathrm{head}_M) W_O \mathrm{\ and\ } \mathrm{head}_i = \mbox{softmax}\Bigg(\frac{Q_i(K_i)^T}{\sqrt{d_k}}\Bigg)V_i,
\end{equation}
where $Q_i = W_Q \times X_i'$, $K_i = W_K \times X_i'$, $V_i = W_V \times X_i'$, $X_i' = [P_i, X_i]$, $M$ is number of attention heads indexed by $i$ and $W_Q, W_K, W_V$, $W_O \in \mathbb{R}^{E \times E}$ are model parameters. 

%\xl{Here you suddenly stopped. What is the relationship before $p$ and $P$?}
\subsection{Token Clustering Module}
\label{sec:loc-glo}
\noindent\textbf{Deep Embedding Clustering (DEC).} The DEC block is designed to cluster ROI into $K$ communities based on an assignment matrix derived from the prompt tokens and node feature embeddings, which are viewed as the centroids for each cluster. To this end, the output token embeddings obtained from the Transformer Encoder Layer (Sec.~\ref{sec:TEL}) $H$ are split into the embedded node features $H_X \in \mathbb{R}^{N \times E}$ and embedded prompt tokens $H_P = [p_1, \dots, p_K] \in \mathbb{R}^{K \times E}$. Each embedding prompt token $p_k$ is normalized to a unit vector by $p_i \gets p_i/\|p_i\|_2$ for $i \in [K]$.
% \begin{equation}
% p_{ki} \leftarrow \frac{p_{ki}}{(\sum_{j=1}^{E}p_{kj}^2)^{0.5}}
% \end{equation}
%The DEC layer (Fig.~\ref{fig:architecture}(b)) first splits the output of the transformer encoder layer $H$ into the embedded node features $H_X \in \mathbb{R}^{N \times E}$ and embedded prompt tokens $H_P \in \mathbb{R}^{K \times E}$. 
The normalized embedded prompt tokens $\bar{H}_P$ are used for clustering ROI tokens and the calculation of the orthogonal loss\footnote[1]{During training, we average each prompt embedding over batch for each step.}. The embedded node features $H_X$ are encoded through a multilayer perceptron (MLP) and combined with the normalized prompt tokens $\bar{H}_P$ with a softmax operation over the community dimension\footnote[2]{The total probabilities of an ROI to be assigned to all the community add up to one.} to generate the soft assignment matrix $A$, derived as
\begin{equation}
A = \mathrm{softmax}(X \times P^T),\ A \in \mathbb{R}^{N \times K}.
\end{equation}
\noindent\textbf{Orthogonal Loss.} To ensure the differences between communities, we aim to enforce the independence and orthogonality of the embedded prompt tokens. We assign each prompt token to a certain community and we aim to maximize the relations between nodes within the same community while minimizing the relations between nodes within different communities. To achieve this, the orthogonal loss is introduced which aims to optimize the cosine similarity of each prompt token, given as
\begin{equation}\label{eq:otho}
\mathcal{L}_\textrm{ortho} = \Big\lVert P \times P^{T} - I_K \Big\rVert_F,
\end{equation}
where $P_k$ is the $k^{\mathrm{th}}$ row of $P$, $I_K$ is identity matrix with dimension $K \times K$ and $\lVert \ \rVert_F$ denotes Frobenius norm for matrix.\\
\noindent\textbf{Token Merging.} The final outputs from the TC module are the merge token representations derived by matrix multiplication of soft assignment matrix and encoded node features given by 
\begin{equation}
H_{\mathrm{out}} = A^T \times H,\ H_{\mathrm{out}} \in \mathbb{R}^{K \times E},
\end{equation}
where $N$ embedded node features are projected to $K$ prompt tokens embedding space. Through this operation, all node features belonging to one cluster are compressed to a single vector, thus reducing the dimensionality. %Additionally, our method requires fewer parameters than \cite{bannadabhavi2023community} (refer to Table~\ref{tab:para}), which helps reduce overfitting when training with small datasets.
% \noindent\textbf{Soft Assignment:} Given prompt tokens $P \in \mathbb{R}^{K \times E}$ optimized by the orthogonal loss and embedded node features $H \in \mathbb{R}^{N \times E}$, the assignment matrix is derived as
% \begin{equation}
% A = \mathrm{softmax}(X \times P^T),\ A \in \mathbb{R}^{N \times K}
% \end{equation}
% and the node representation is given by
% \begin{equation}
% H_{\mathrm{out}} = A^T \times H,\ H_{\mathrm{out}} \in \mathbb{R}^{K \times E}
% \end{equation}
% where $N$ embedded node features are projected to $K$ prompt tokens embedding space and all node features belonging to one cluster are compressed to a single vector.

\subsection{Graph Readout Layer}

The final step involves mapping node features embedding onto the prompt tokens embedding space to obtain a high-level representation of the brain graph. We implement a modified OCRead layer \cite{kan2022bnt} for aggregating the learnt node features embedding. The embedding dimension of node representations $Z$ is reduced to $E^\prime$ and then feature dimension and feature embedding dimension are combined to a single vector with shape $(K \times E^\prime)$. $Z$ is then flattened to shape $(K \times E^\prime)$ and passed to an MLP for a graph-level prediction. The whole process is supervised with Cross-Entropy (CE) loss.

%% file: section/03-experiments.tex
\section{Experiments}

\subsection{Datasets and Experimental Settings}

 %This dataset provides an opportunity to study brain structure, function, and connectivity, and to explore their relationships to cognitive, behavioral, and health outcomes. Each sample is labeled as either male or female.

\noindent\textbf{ABIDE Dataset.} encompasses resting-state functional MRI (rs-fMRI) data sourced from 17 international sites~\cite{di2014autism}, processed using the Configurable Pipeline for the Analysis of Connectomes (CPAC). This processing includes band-pass filtering (0.01 - 0.1 Hz) without the application of global signal regression. The dataset undergoes parcellation according to the Craddock 200 atlas~\cite{Craddock2012}, resulting in $N=200$ Regions of Interest (ROIs). %These ROIs are further classified into eight functional communities as per the assignments detailed in \cite{yeo2011organization}: cerebellum and subcortical structures (CS \& SB), visual network (V), somatomotor network (SMN), dorsal attention network (DAN), ventral attention network (VAN), limbic network (L), frontoparietal network (FPN), and default mode network (DMN). Comprising rs-fMRI data from 1009 participants.
The ABIDE dataset includes 493 healthy controls (HC) and 516 individuals diagnosed with Autism Spectrum Disorder (ASD). The dataset's analytical approach involves the computation of Pearson correlation matrices from the mean time series of the ROIs, with each participant's data labeled as either ASD or HC.

\noindent\textbf{HCP Dataset.} in our evaluation refers to a subset of the original Human Connectome Project (HCP) dataset from 688 healthy adults, comprised of 294 males and 394 females~\cite{harms2018extending}. We perform gender classification using their fMRI data registered to Shen268 altas~\cite{shen2013groupwise}, resulting in the number of ROIs $N=268$.

\noindent\textbf{Experimental Settings.} %We trained for the binary classification task on both datasets. 
We set the transformer encoder layer embedding dimension $E=512$, $E^\prime=128$, and number of heads $M=8$ as default. We vary the number of communities $K \in \{4,8,11\}$. %All models are implemented in PyTorch 1.12 and trained on NVIDIA RTX 3090 with 24GB memory. 
We used the Adam optimizer with an initial learning rate of $10^{-4}$ and a weight decay of $10^{-4}$. We set the learning rate of each parameter group using a cosine annealing schedule with warm restart\cite{loshchilov2016sgdr}, where the number of iterations for the first restart is 10 and the minimum learning rate is $10^{-5}$. The batch size is 64 and models are trained for 50 epochs. We use the area under the receiver operating characteristic (AUROC), accuracy, sensitivity, and specificity on the test set to evaluate performance. The epoch with the highest AUROC performance on the validation set is used for performance comparison on the test set. Following the setting in \cite{bannadabhavi2023community}, to prevent imbalance in the distribution of target classes, we use stratified sampling strategy\cite{neyman1992two} for train-validation-test data split, resulting in the same train, validation and test data with split ratio 70:10:20. The mean and standard deviation values of the metrics of five random runs are reported for performance comparison.

\subsection{Quantitative and Qualitative Results}

\noindent\textbf{Quantitative Analysis.} We evaluated our proposed model - \ours{}, against both transformer-based models and popular deep learning architectures designed for brain network analysis. Among the transformer-based models, we compare with BrainNetTF, which employs a soft clustering strategy in its readout phase for Regions of Interest (ROIs), and Com-BrainTF, a model that achieves state-of-the-art (SOTA) performance by leveraging predetermined clusters in the transformer encoder. Additionally, we compare the performance of our model against well-established deep-learning models tailored for brain networks. BrainNetCNN represents a category of models that utilize a fixed network structure with specialized graph neural network (GNN) approaches, while FBNETGNN exemplifies models that feature a learnable network structure, becoming the current leading edge in adaptable GNNs. As shown in Table 1, \ours{} (K=11) achieves the best performance across the AUROC, Accuracy and Specificity on the ABIDE dataset and the best performance across AUROC, Accuracy and Sensitivity on the HCP dataset. \ours{} stands out by not confining itself to the limitations of predefined community structures. Instead, it dynamically learns and captures the complex relationships within and between functional communities, offering a representation of brain connectivity that aligns closely with the inherent organizational patterns of the brain. We also note that \ours{} consistently achieves better results on overall metrics on HCP with different $K$. However, it suffers from dropped performance when $K=4$ or $8$ on ABIDE, from which we observe the averaged performance is affected by the very low performance obtained from one run. Still, our performance in these settings is better than several baselines, such as BrainNetTF~\cite{kan2022bnt} and FBNETGNN~\cite{Fbnetgen}.
%It is also demonstrated in Table 1 that a larger number of prompt tokens outputs better results in terms of classification metrics. However, a larger number of prompt tokens leads to larger model parameter numbers as shown in \ref{tab:para}.

\begin{table}[!t]
\caption{Quantitative results for classification compared with baselines (mean $\pm$ standard deviation) on ABIDE and HCP dataset.}\label{tab:quantitative}
\resizebox{\textwidth}{!}{%
\begin{tabular}{lllllllll}
\hline
\multirow{2}{*}{Method} & \multicolumn{4}{l}{ABIDE} & \multicolumn{4}{l}{HCP} \\ \cline{2-9} 
 & AUROC & Accuracy & Sensitivity & Specificity & AUROC & Accuracy & Sensitivity & Specificity \\ \hline
FBNETGNN\cite{Fbnetgen} & $73.8_{\pm 3.4}$ & $65.2_{\pm 2.0}$ & $71.4_{\pm 5.3}$ & $59.0_{\pm 8.3}$ & $84.7_{\pm 4.2}$ & $76.8_{\pm 4.3}$ & $82.1_{\pm 5.9}$ & $69.7_{\pm 8.6}$ \\
BrainNetCNN\cite{KAWAHARA20171038} & $75.0_{\pm 3.4}$ & $69.0_{\pm 5.1}$ & $77.8_{\pm 5.6}$ & $60.9_{\pm 7.6}$ & $85.5_{\pm 3.2}$ & $78.5_{\pm 3.0}$ & $80.5_{\pm 4.8}$ & $75.9_{\pm 7.2}$ \\
BrainNetTF\cite{kan2022bnt} & $75.9_{\pm 3.8}$ & $66.0_{\pm 3.2}$ & $68.4_{\pm 14.6}$ & $62.0_{\pm 16.3}$ & $88.2_{\pm 1.9}$ & $77.9_{\pm 3.1}$ & $82.6_{\pm 4.4}$ & $71.7_{\pm 3.4}$ \\
Com-BrainTF\cite{bannadabhavi2023community} & $77.3_{\pm 3.9}$ & $65.6_{\pm 6.3}$ & $\mathbf{83.9_{\pm 7.0}}$ & $47.6_{\pm 17.5}$ & $88.7_{\pm 2.3}$ & $79.1_{\pm 4.1}$ & $80.5_{\pm 8.5}$ & $\mathbf{77.2_{\pm 9.4}}$ \\
\ours{} ($K=4$)& $76.2_{\pm 3.5}$ & $65.8_{\pm 4.5}$ & $71.5_{\pm 17.8}$ & $62.7_{\pm 20.2}$ & $88.1_{\pm 2.7}$ & $81.8_{\pm 1.5}$ & $86.7_{\pm 4.7}$ & $75.2_{\pm 9.1}$ \\
\ours{} ($K=8$)& $74.0_{\pm 3.3}$ & $64.2_{\pm 7.7}$ & $73.4_{\pm 9.1}$ & $57.5_{\pm 18.3}$ & $89.5_{\pm 1.9}$ & $79.4_{\pm 2.6}$ & $83.1_{\pm 6.0}$ & $74.5_{\pm 3.5}$ \\
\ours{} ($K=11$) & $\mathbf{77.7_{\pm 2.0}}$ & $\mathbf{69.4_{\pm 2.9}}$ & $69.1_{\pm 8.6}$ & $\mathbf{70.1_{\pm 3.9}}$ & $\mathbf{89.9_{\pm 2.1}}$ & $\mathbf{80.9_{\pm 2.5}}$ & $\mathbf{86.7_{\pm 3.8}}$ & $73.1_{\pm 4.0}$ \\ \hline
\end{tabular}
}
%\vspace{-2mm}
\end{table}

\begin{figure}[!t]
\includegraphics[width=1.0\textwidth]{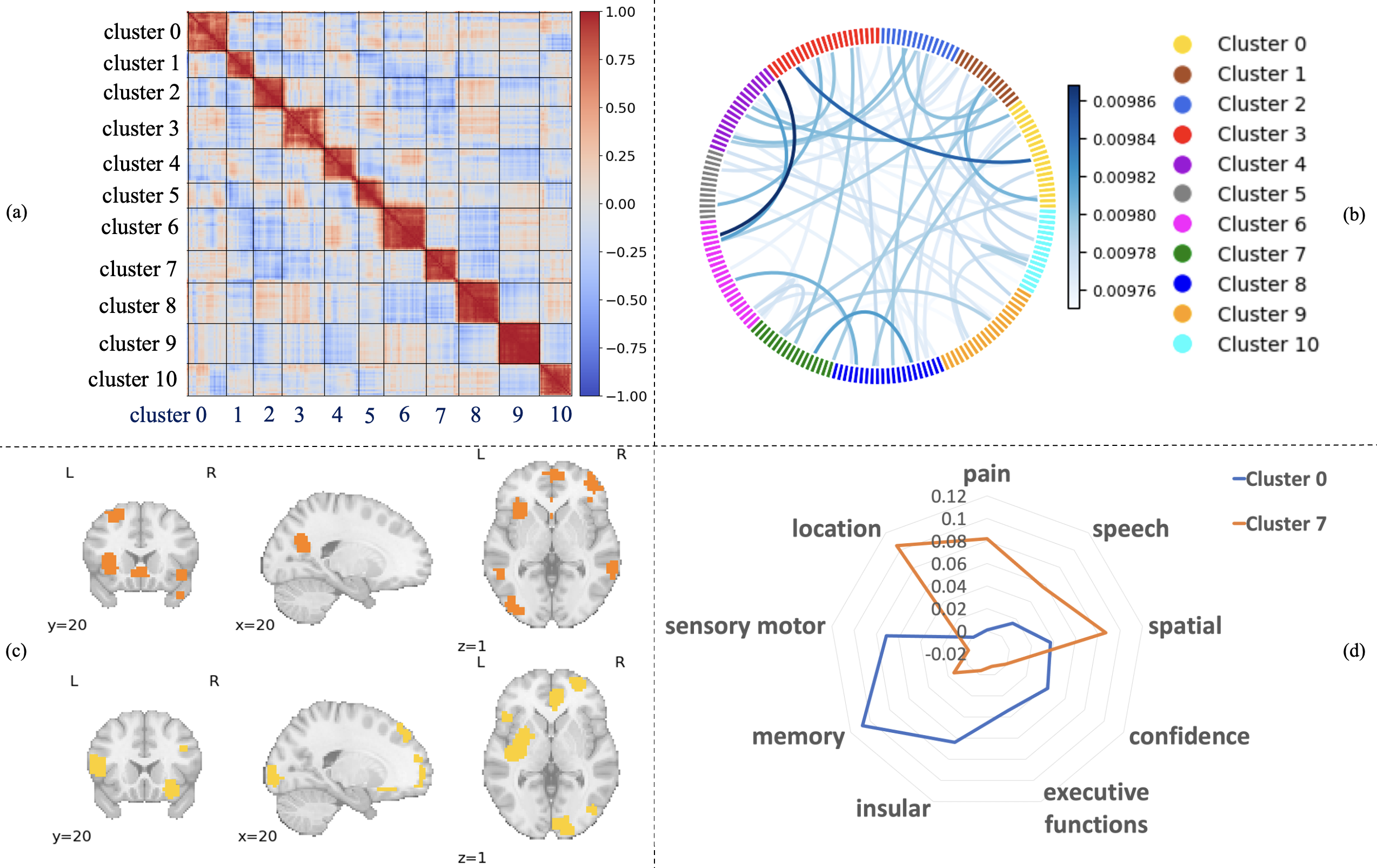}
\caption{Qualitative results for communities clustering. (a) Assignment matrix $A$ depicting the relationship between ROIs. (b) Chord diagram showcasing the top 0.5\% strongest connections from the token-token attention scores. (c) The ROIs in salient clusters 0 and 7 (d) Radar chart representing the correlation between functional clusters and cognitive keywords decoded by Neurosynth.}
\label{fig:qualitative}
\vspace{-4mm}
\end{figure}

We present the ablation study on orthogonal loss (Eq.~\eqref{eq:otho}) in Appendix.

\noindent\textbf{Qualitative Analysis.} Our qualitative analysis on ASD dataset with $K=11$ examines 1) the soft-assignment matrix $A$, reflecting the similarity between ROIs. We average $A$ over correctly classified test data and normalizing along the embedding dimension, 2) the $N \times N$ token-token attention scores $S_{\rm TT}$, and $K\times N$ prompt-token attention scores $S_{\rm PT}$ in the Transformer Encoder Layer (Sec.~\ref{sec:TEL}). 
To make it easier to visualize the assignment matrix $A$ in Fig.~\ref{fig:qualitative}(a), we organize the ROI by grouping them according to the hard cluster index (taking the argmax across the cluster number dimension for $A$). We observe significantly stronger intra-cluster similarities than inter-cluster similarities, demonstrating the effectiveness of the proposed token clustering and emerging operation. We also visualize the attention scores $S$ averaged over all the correctly classified testing subjects in the Chord diagram of Fig.~\ref{fig:qualitative}(b), which depicts the inter-cluster and intra-cluster interaction both occur towards making classification decision. 
%the Assignment Matrix, derived by extracting the soft-assignment matrix pre-softmax layer for correctly classified test data and normalizing along the embedding dimension. The matrix captures the level of attention that each ROI directs towards others. The visibly brighter diagonal underscores a significant intra-cluster similarity, signifying the proposed toke emerging operation. 
To interpret the important clusters associated with the task, we look at $S_{\rm PT}$ as we determine the prediction on merged tokens with prompts as centroid and average attention scores in $S_{\rm PT}$ for each cluster over its ROIs. The averaged attention scores for cluster 0 and cluster 7 stand out, implying a potentially pivotal role in ASD-related brain functions.
Fig.~\ref{fig:qualitative}(c) displays the ROIs for clusters 0 and 7. We further conduct meta-analysis on these clusters by employing the Neurosynth database~\cite{yarkoni2011large}, elucidating the functional significance of the identified clusters. As shown in Fig.~\ref{fig:qualitative}(d), Cluster 7 exhibits a stronger correlation with cognitive keywords associated with location, pain, speech, and spatial functions. In contrast, Cluster 0 has a marked association with terms related to memory, insular, executive functions, and confidence, hinting at cognitive domains that might be distinctly affected in conditions such as ASD~\cite{hodges2020autism}. The analysis for HCP is available in the Appendix.